\documentclass[]{spie}  %>>> use for US letter paper
\usepackage{makecell}

\usepackage{amsmath,amsfonts,amssymb}
\usepackage{graphicx}
\usepackage[colorlinks=true, allcolors=blue]{hyperref}
\usepackage[numbers,sort]{natbib}
\title{HighSpec: A High-Resolution Spectrograph for the MAST Telescope Array}

\author[a]{Yahel Sofer Rimalt}
\author[a]{Sagi Ben-Ami}
\author[a]{Eran Ofek}
\author[a]{Na'ama Hallakoun}
\author[a]{Ido Irani}
\author[a]{Oren Ironi}
\author[b]{Jani Achren}
\author[a]{Alex Bichkovsky}
\author[a]{Arie Blumenzweig}
\author[a]{Ofir Hershko}
\author[c,d]{Hanindyo Kuncarayakti}
\author[c]{Seppo Mattila}
\author[e]{Tsevi Mazeh}
\author[a]{Gleb Mikhnevich}
\author[a]{David Polishook}
\author[a]{Ofer Yaron}

\affil[a]{Weizmann Institute of Science, 234 Herzl St., Rehovot, 7610001, Israel}    
\affil[b]{Incident Angle Oy, Capsiankatu 4 A 29, FI-20320 Turku, Finland}
\affil[c]{Tuorla observatory, Department of Physics and Astronomy, University of Turku, V\"ais\"al\"antie 20, 21500 Piikki\"o, Finland}
\affil[d]{Finnish Centre for Astronomy with ESO (FINCA), FI-20014 University of Turku, Finland}
\affil[e]{Tel-Aviv University, 55 Chaim Levanon St., Tel-Aviv, 6997801, Israel}

\authorinfo{Send correspondence to yahel.sofer-rimalt@weizmann.ac.il}

\begin{document} 
\maketitle

\begin{abstract}

We present the updated design of HighSpec, a high-resolution $\mathcal{R}  \sim  20,000$ spectrograph designed for the Multi Aperture Spectroscopic Telescope (MAST). HighSpec offers three observing modes centered at the Ca\,II H\&K, Mg\,b triplet, and H$\alpha$ lines. Each mode is supported by a highly optimized ion-etched grating, contributing to an instrument exceptional peak efficiency of $\gtrsim85\%$ for the two latter bands ($\gtrsim55\%$ for the Ca\,II H\&K band). 
Optimizing throughput over wavelength coverage ($\Delta \lambda=10-17\, $nm), HighSpec enables the precise measurement of spectral lines from faint targets. This approach is especially relevant for stellar object studies, specifically of WDs, which are intrinsically faint and have few spectroscopic lines. Each observing mode was tailored to target spectral features essential for WD research. 
Its integration with MAST, an array of 20 custom-designed telescopes that can function as a single large telescope (equivalent to a $2.7\,$m telescope in collecting area) or multiplexing over the entire sky, provides unique adaptability for extensive and effective spectroscopic campaigns. Currently in its final assembly and testing stages, HighSpec's on-sky commissioning is scheduled for 2025.

\end{abstract}


% -----------------------------------------------------
\section{INTRODUCTION}
\label{sec:intro}  % \label{} allows reference to this section
% -----------------------------------------------------
High-resolution spectroscopy at visible wavelengths is one of the key tools in modern astrophysics. Balancing resolution, wavelength coverage, and instrument efficiency is critical and can significantly improve our observing capabilities. Most high-resolution instruments are Echelle spectrographs, which cover a relatively broad spectral range and generally have low efficiencies. This inefficiency is mainly due to the use of an Echelle grating, a cross-disperser, and a large number of optical components. Because of the inherently low signal expected per resolution element at high dispersion and the overall low efficiency, these instruments are usually coupled to medium to large telescopes. As a result, high-resolution spectroscopic studies are often limited to relatively bright targets, require competitive observation time on one of the available large telescopes, and are restricted by a close set of trade-off relations between the observing properties offered by each of the instruments. 

Developing instruments tailored for particular science cases and facilities can greatly enhance the sensitivity of the measurement and boost the scientific output. This approach is particularly beneficial for the study of white dwarfs (WDs), which are characterized by their low brightness and have a distinct spectroscopic signature, as due to their high surface gravity their spectra are typically dominated by a few broad lines. Consequently, there is little need for large bandpass coverage or very high resolution in most WD spectroscopic studies, making specialized instruments particularly advantageous. 

Here, we present HighSpec -- a high-resolution $\mathcal{R}  \sim  20,000$ spectrograph featuring three observing modes, specifically optimized for the study of WDs. These include a dedicated mode for radial velocity (RV) measurements, utilizing the non-local thermal equilibrium (NLTE) core of the H$\alpha$ line, and two modes that were designed to detect and analyze contamination signatures of calcium (Ca\,II H\&K) and magnesium (Mg\,b triplet) in WD atmospheres. The design of HighSpec reflects our choice to maximize the throughput, reaching an instrument peak efficiency of 88\% at the cost of utilizing narrow band-passes and abandoning simultaneous wavelength coverage. Table~\ref{tab:obsModes} includes a summary of the observing modes.

HighSpec is a fiber-fed spectrograph designed for MAST, an array of twenty 61\,cm aperture telescopes. The incoming fibers from the telescope array comprise a long slit on the object plane and are imaged to separate traces on the detector. %To reduce the penalty of the impact of multiple traces from multiple independent telescopes, we employ an electron-multiplying CCD (EMCCD) as the detector.
The integrated system, comprising both MAST and HighSpec, supports diverse operational modes. The design allows HighSpec to function either as a single-object spectrograph coupled to a $\sim2.7\,$m telescope or as a multi-object spectrograph for small $\sim60\,$cm telescopes that is not confined to a single field of view, enabling extensive multiplexing capabilities. 
A system-level view of the MAST telescope-to spectrum is presented in Figure~\ref{fig:systemDiagram}.
Given its automated rapid switching option between observing modes, HighSpec is ideally suited for comprehensive and detailed surveys of WDs, able to observe more than 6,500 known WDs \cite{gentile2021catalogue}.

The paper is organized into three main sections. The first section provides an overview of the MAST telescope array, emphasizing its unique flexibility property. The second section offers a detailed description of HighSpec design, the technologies it adopts, and expected performance. The final section discusses the WD case studies that influenced the design of HighSpec.

\begin{table}[ht]
    \caption{HighSpec observing modes. Instrument efficiency refers to slit-to-detector. End-to-end efficiency represents the total expected efficiency, sky-to-detector.} 
    \label{tab:obsModes}
    \begin{center}       
        \begin{tabular}{|l|c|c|c|c|}
            \hline
            \makecell{Observing\\ Mode} & \makecell{$\lambda_0$ \\ $[$nm]} & \makecell{$\Delta \lambda$ \\ $[$nm]} & \makecell{Instrument \\ Efficiency [\%] } & \makecell{End-to-End \\ Efficiency[\%]} \\
            \hline \hline
            Ca\,II H\&K & $395$ & $10.5$ & 55 & 31 \\
            \hline
            Mg\,b  & $518$ & $13.8$ & 88 & 55 \\
            \hline
            H$\alpha$ & $656$ & $17.5$ & 85 & 55\\
            \hline
        \end{tabular}
    \end{center}
\end{table}

% -----------------------------------------------------
\section{THE TELESCOPE ARRAY -- MAST}
% -----------------------------------------------------
The Multi Aperture Spectroscopic Telescope (MAST) is a spectroscopic array of twenty custom-designed $61\,$cm Newtonian telescopes, currently under construction at the newly established Weizmann Astronomical Observatory (WAO) in Ne'ot Smadar, southern Israel. Manufactured by PlaneWave Instruments, each telescope in the array features a prime mirror design operating at F/3, utilizing a parabolic mirror. It has a collective aperture equivalent to a single $2.68\,$m telescope, providing significant observational power at approximately 10\% of the cost of a similar-sized single telescope \cite{ofek2020seeing}.

Each telescope is equipped with two optical fibers, one for the science target and the second for the surrounding sky. % The plate scale of MAST, working at F/3, gives a projected sky aperture of 3.48\,arcsec for our fibers. 
The fibers termini from all telescopes comprise a $\sim\,5\,$mm-long quasi-slit unit that can be plugged into one of two instruments on MAST optical table: (i) DeepSpec - a low spectral resolution $\mathcal{R}  \sim  650$ broadband spectrograph \cite{Ironi2024DeepSpec}; (ii) HighSpec: a high-resolution narrow bandpass spectrograph. 

The two instruments image the light traces from different telescopes separately on the detector.
As each telescope will have an independent guiding and tracking system, this arrangement ensures that each telescope can function independently, providing flexibility in observing strategies.
Therefore, MAST is not confined to a single target or a single field of view, enabling extensive multiplexing capabilities. This design allows HighSpec to function either as a multi-object spectrograph, permits multiplexed operations across different sky regions, or a single-object spectrograph, making it exceptionally suited for large population surveys.

%The telescopes' mounts (PlanWave L600) offer a tracking precision $\lesssim 1$ arcsec per minute with wind speed $<10$ km/s.

HighSpec is expected to utilize half of the site's clear nights ($>300$ night/year), for $>3$ years.

\begin{figure} [ht]
   \begin{center}
       \begin{tabular}{c} %% tabular useful for creating an array of images 
        \includegraphics[height=3.3cm]{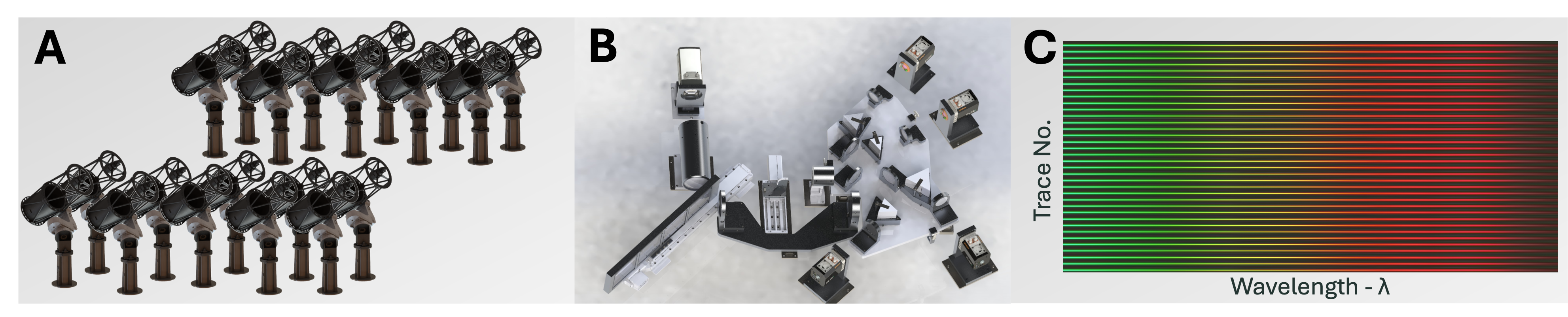}
       \end{tabular}
   \end{center}
   \caption[example] 
   { \label{fig:systemDiagram} 
   MAST System Overview: A) The array features twenty $61\,$cm Newtonian telescopes organized into two sub-arrays of ten, each housed in a dedicated enclosure. Optical fibers transport light for both target and sky from each telescope to the instrument room. B) The MAST optical table, accommodating both HighSpec and DeepSpec spectrographs. C) The final signal is distributed across multiple individual traces.}
\end{figure}

A top-level illustration of the system is presented in Figure~\ref{fig:systemDiagram}. Table \ref{tab:tel_params} summarizes some basic properties of the WAO observatory and the custom PlaneWave telescopes used for MAST.

\begin{table}[ht]
\caption{Telescope and observatory parameters} 
\label{tab:tel_params}
\begin{center}       
\begin{tabular}{|l|c|}

\hline
Property & Value \\
\hline \hline
Single telescope effective diameter [mm] & 580 \\
\hline
Focal length [mm] & 1830   \\
\hline
Number of telescopes & 20 \\
\hline
Telescopes per mount & 1 \\
\hline
Dark time sky brightness [$\rm mag / arcsec^2$] & 20.5  \\
\hline
Telescope mirror reflectivity [\%]& $>90$ \\
\hline
Median seeing [arcsec] & 1.35  \\
\hline
Fiber core size [$\mu m$] & 30 \\ 
\hline
Fiber end-to-end throughput (reflectivity and FRD)[$\%$] & 86\\
\hline
Location & Ne'ot Smadar, Israel \\
\hline
Longitude (WGS84) [deg] & 35.0407331 E \\
\hline
Latitude (WGS84) [deg] & 30.0529838 N \\
\hline
Height (WGS84) [m] & 415 \\
\hline

\end{tabular}
\end{center}
\end{table}

\begin{figure} [ht]
   \begin{center}
       \begin{tabular}{c} %% tabular useful for creating an array of images 
        \includegraphics[height=7cm]{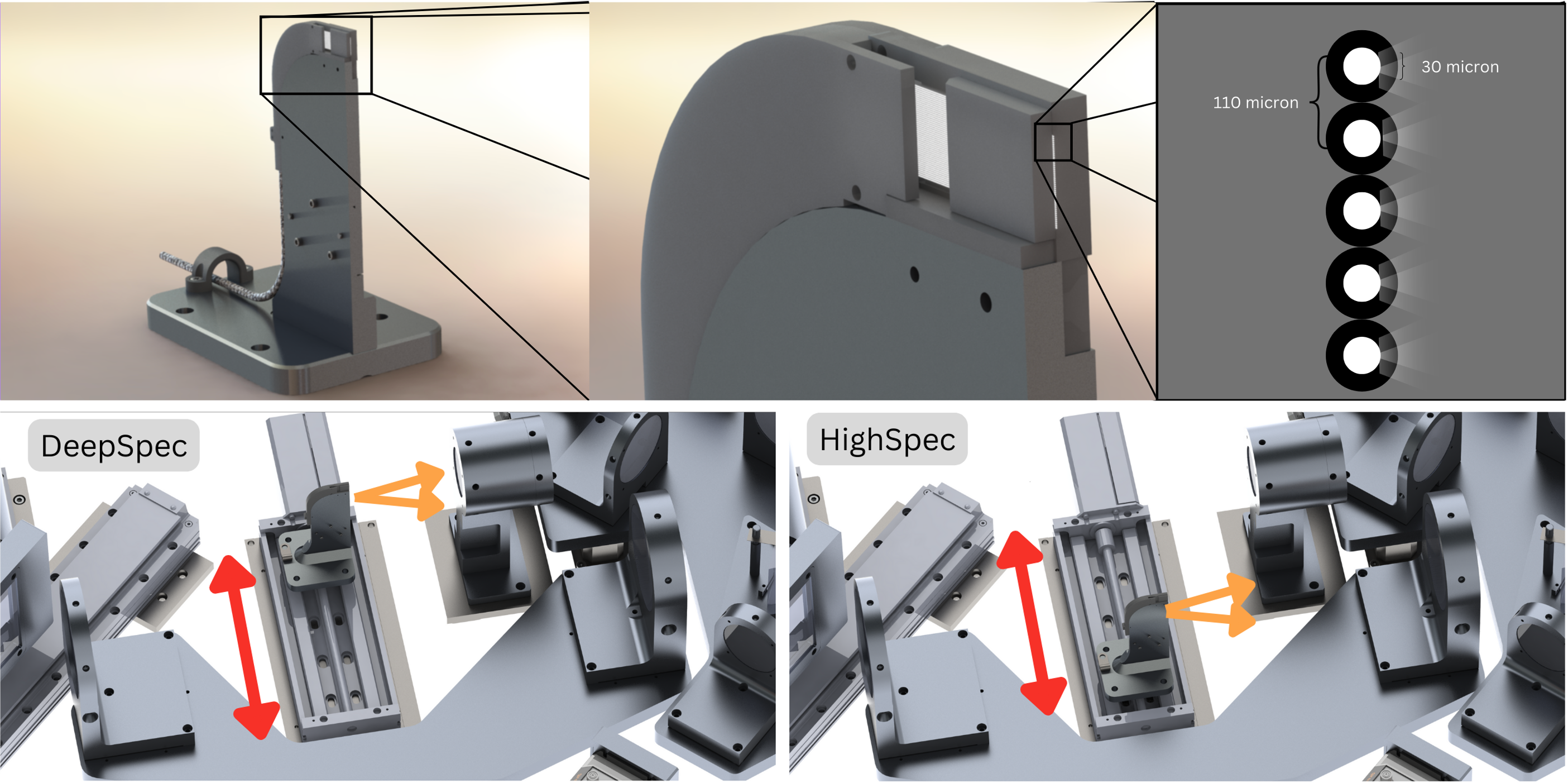}
       \end{tabular}
   \end{center}
   \caption[example] 
   { \label{fig:fiberSlit} 
    The fiber slit. Top Panel: Each fiber core has a diameter of $30\mu$\,m and a center-to-center distance of $110\mu m$. The 45 fibers in the assembly have a total length of $\sim5$\,mm. Bottom Panel: The unit is mounted on a linear stage to transition between DeepSpec and HighSpec.
}
\end{figure}

% -----------------------------------------------------
\subsection{MAST fibers}
\label{sec:MASTfibers}
% -----------------------------------------------------
Each of the MAST telescopes is equipped with two Few Modes (FM) fused silica, broad-spectrum, $30\,\mu$m core optical fibers with a $\sim 15$m length from Polymicro. 
The use of FM optical fibers increases the fiber throughput when compared to the use of a single mode fiber for a non-Gaussian and imperfect beam as expected in the case of an astronomical telescope observing under sub-optimal seeing conditions, reaching throughput of over $ 86\%$ as verified in the lab. The projected fiber core angle on sky is $3.44\,$arcsec, and the projected separation between the target and sky fibers is about 10\,arcmin. As the median seeing at the site measured at Zenith over the course of a year is $\sim$1.35 arcsec \cite{ofek2023large}, and the guiding precision of the mount is $\lesssim1$ arcsec, after an adequate system performance is achieved we plan to upgrade the fibers to $25\,\mu$m core fibers ($2.85$\,arcsec on the sky), which will result in a $20\%$ increase in the spectral resolution.

The 40 fibers from the MAST telescopes (20 for science and 20 for sky) are combined with five additional fibers from the calibration system (three active and two spares) into a 45-fibers long-slit. The unit is mounted on a Zaber LRQ300AL linear stage for transitioning between DeepSpec and HighSpec so that the fiber facets can be positioned in each of the two spectrographs collimator focal points, see Figure~\ref{fig:fiberSlit} bottom panel.

%The 40 fibers from the MAST telescopes (20 for science and 20 for sky), as well as the five fibers from the calibration system, combined into a 45-fibers long-slit, are mounted on a Zaber LRQ300AL linear stage for transitioning between DeepSpec and HighSpec so that the fiber facets can be positions in each of the two spectrographs collimator focal point, see Figure~\ref{fig:fiberSlit} bottom panel.

% -----------------------------------------------------
\subsection{Injection Unit - Flexure Indicating Fiber Feed Assembly - FIFFA}
\begin{figure} [ht]
   \begin{center}
       \begin{tabular}{c} %% tabular useful for creating an array of images 
        \includegraphics[height=6cm]{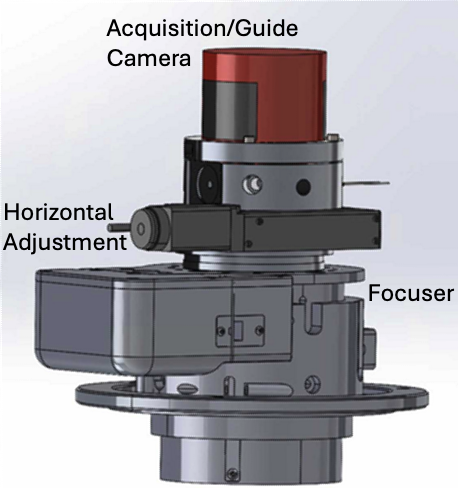}
       \end{tabular}
   \end{center}
   \caption[example] 
   { \label{fig:FIFFA} 
    The guide and fiber injection unit --- FIFFA. The unit supports telescope guiding on field stars and allows target acquisition and telescope focusing}
\end{figure}

% -----------------------------------------------------
Each telescope is equipped with a guide and injection unit - a Flexure Indicating Fiber Feed Assembly (FIFFA). The unit, designed and built by the Weizmann Institute of Science Astrophysical Instrumentation group, allows us to monitor the injection of light from a specific target into an optical fiber. 
Within each FIFFA unit, a 45-degree fold mirror with a high-efficiency dielectric coating (R$\gtrsim98.5\,\%$) directs the on-axis field to the optical fiber ports. The surrounding field, divided into two sub-areas, each approximately $0.4^\circ\times0.2 ^\circ$  (with $\leq50\,\%$ vignetting and subject to coma aberrations), is imaged onto a guide camera. This setup facilitates guiding on field stars, operating in a closed loop (~0.1 Hz) with the PlaneWave L550 mount to stabilize the star's image on the fiber face.
Mounted on a linear motorized stage, FIFFA facilitates horizontal adjustment relative to the telescope for target acquisition and focusing. Before observations, the unit is moved to align the target image with the telescope’s central focus. FIFFA then precisely adjusts to shift the target image from the alignment position to the folding mirror, ensuring accurate delivery to the optical fiber.
As both the MAST array and its instruments DeepSpec and HighSpec operate at F/3, we do not implement an additional relay system to speed up the beam and reduce expected FRD. Having a single element in the target’s optical path, improves the design's simplicity and significantly boosts the unit's efficiency.
A CAD model of the FIFFA unit is shown in Figure~\ref{fig:FIFFA}.

% -----------------------------------------------------
\subsection{Calibration}
\label{sec:calib}
% -----------------------------------------------------
%High-precision, stabilized spectrographs require regular wavelength calibration due to environmental changes such as pressure and temperature fluctuations, which can affect the spectral response both in the short and long terms. 
%
To facilitate wavelength calibration, the quasi-slit is designed to accommodate five additional calibration fibers (three active and two spared) spaced among the science/sky fibers. Each fiber delivers similar intensity and spectral profiles to ensure consistent, identical traces.
The MAST calibration unit employs two light sources: A Quartz Tungsten-Halogen (QTH) lamp serves as a spectral flat calibration source, and a Thorium-Argon (ThAr) hollow cathode lamp is used to generate arc spectra for wavelength calibration. 
The system is designed for easy switching between the two light sources, enabling efficient transitions.
For further details, please refer to Kuncarayakti et al., 2024 \cite{Kuncarayakti2024Calib}.

\begin{figure} [ht!]
   \begin{center}
   \begin{tabular}{c} %% tabular useful for creating an array of images 
   \includegraphics[height=5cm]{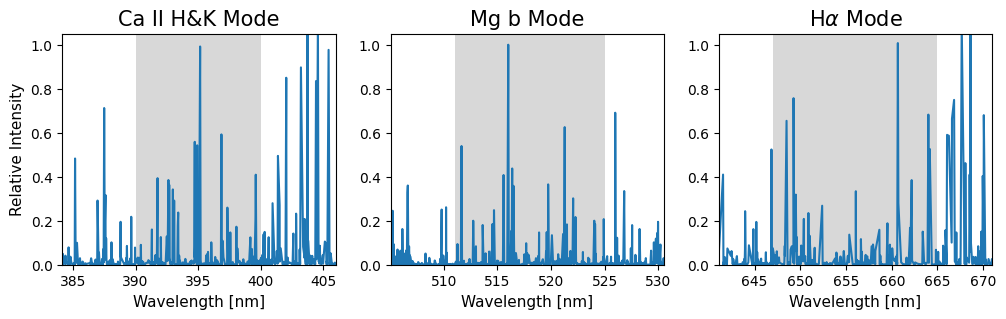}
   \end{tabular}
   \end{center}
   \caption[example] 
%>>>> use \label inside caption to get Fig. number with \ref{}
   { \label{fig:calibration}  Available emission lines of the Thorium-Argon (ThAr) calibration source, as referenced from Lovis and Pepe (2007) \cite{lovis2007new}. The graph shows the relative intensity of the emission lines across the three observing modes of HighSpec. The gray areas highlight the specific wavelength bands covered by HighSpec in each mode, illustrating how the emission lines are distributed within these ranges.}
\end{figure}

%%%%%%
As HighSpec is a narrow-band spectrograph, it is crucial to guarantee an adequate number of calibration lines per mode for an accurate wavelength solution. This is particularly crucial for the H$\alpha$ mode, which is dedicated for RV measurements \footnote{The instrument is not designed to conduct precise RV measurements. We aim to achieve a precision of a few km/s. For more details, please refer to section~\ref{sec:RV}.}. 
Figure \ref{fig:calibration} presents the available emission lines from the Th-Ar source across each of the observing modes, taken from the list of Lovis and Pepe, 2007 \cite{lovis2007new}. It is evident from the figure that there are sufficient lines available for RV calibration in all three modes, ensuring robust wavelength accuracy and instrument stability.

% -----------------------------------------------------
\section{THE INSTRUMENT}
% -----------------------------------------------------
%Overview +table of summary 
%Table \ref{tab:HighSpec_overview} summarizes some basic properties of the HighSpec spectrograph. 

%Table \ref{tab:obsModes} summarizes HighSpec observing modes.

%Fig. \ref{fig:HighSpec_Layout} shows the optical layout of the system.

% -----------------------------------------------------
\subsection{Optical Design}
\label{sec:opticalDesign}
% -----------------------------------------------------

\begin{figure} [ht]
   \begin{center}
       \begin{tabular}{c} %% tabular useful for creating an array of images 
        \includegraphics[height=7cm]{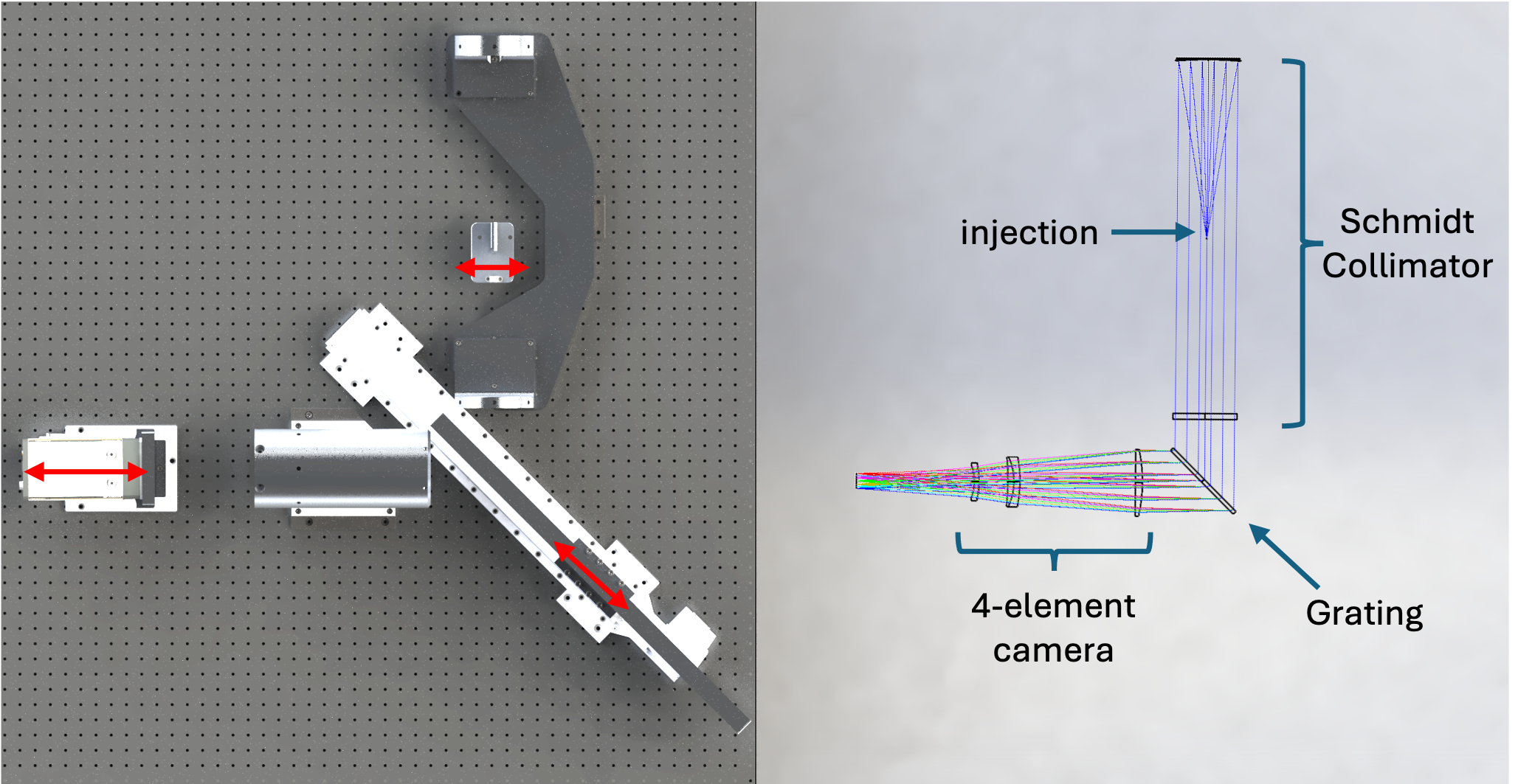}
       \end{tabular}
   \end{center}
   \caption[example] 
  % { \label{fig:HighSpec_Layout}
   { \label{fig:HighSpec_Layout} HighSpec Layout. Left Panel: CAD model of HighSpec, featuring three movable components, each coupled to a motorized stage, with their motion directions indicated by red arrows. These movements facilitate: switching between HighSpec and DeepSpec on the MAST optical table (injection unit), changing between gratings and moving the detector to the focal plane of each mode. Right Panel: ZEMAX model, incorporating ray tracing to demonstrate the light paths through the spectrograph visually. }
\end{figure}

The optical path of HighSpec incorporates a single arm, designed to efficiently switch between the three distinct spectral bands, see Figure~\ref{fig:HighSpec_Layout} for the system layout. The first element in the optical train is a Schmidt camera, which accepts an F/3 beam emerging from the quasi-slit fiber terminus and generates a 100 mm diameter collimated beam. The collimator comprises a spherical mirror and an aspheric corrector located at the realized beam pupil.

To fulfill the Nyquist sampling criteria per resolution element and allow for potential future upgrades to the fibers (see Section~\ref{sec:MASTfibers}), our initial design requires that each resolution element is sampled by three pixels. Therefore, $\Delta l = 3 \times \Delta_{pixel} = 3 \times 16\,\mu m = 48\,\mu m$, based on our detector's pixel size (detailed in Table \ref{tab:detector_parameters}). Consequently, the system's magnification is chosen to be $M = \frac{\Delta l}{d} = \frac{48}{30} = 1.6$.

Following dispersion by a selected ion-etched grating (see section \ref{sec:grating}), a 4-element camera situated 100 mm from the grating mid-plane and operating at F/4.8, images the dispersed beam. The camera's first lens includes an aspheric surface.

\begin{figure} [ht]
   \begin{center}
       \begin{tabular}{c} %% tabular useful for creating an array of images 
        \includegraphics[height=5cm]{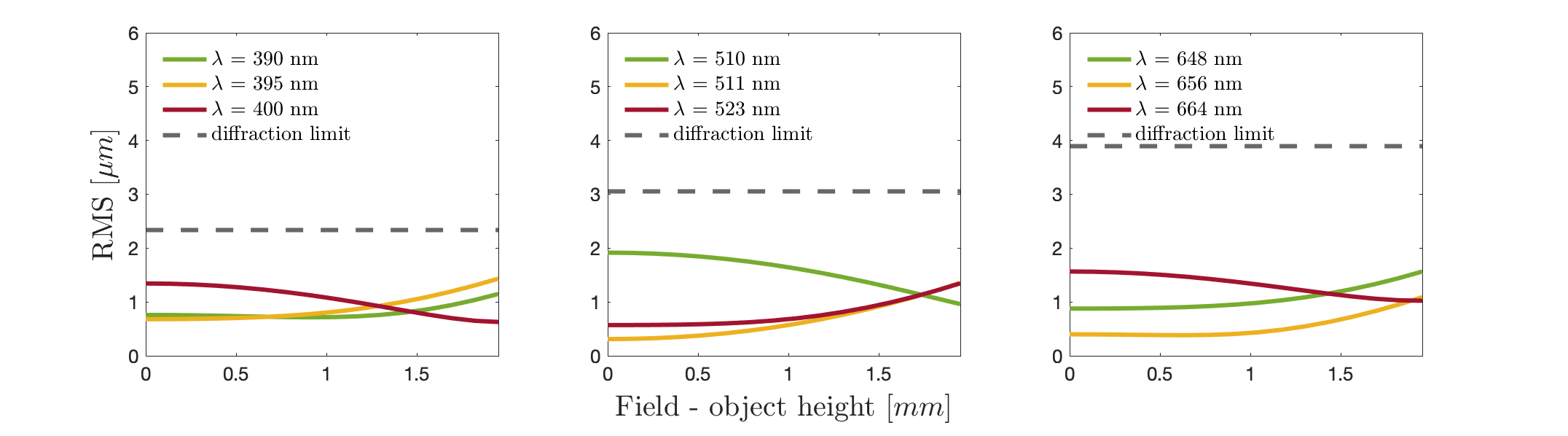}
       \end{tabular}
   \end{center}
   \caption[example] 
   { \label{fig:RMS} HighSpec nominal optical performance ---  spot size RMS.} 
\end{figure}

The nominal spot RMS sizes for all three observing modes and all fields (fibers) are presented in Figure~\ref{fig:RMS}, with values demonstrating performance below the diffraction limit. Despite the simplicity of our optical design, this level of performance is achievable due to the narrow bandwidth of each observation; the significant yet linear longitudinal chromatic aberration when switching between one bandpass and another is effectively mitigated by positioning the detector on a motorized linear stage and conducting focusing adjustment for each observing mode, complemented. 

All the glasses used in the optical path are I-LINE glasses, ensuring high transmission efficiency below $\lambda=400$\,nm. The overall optics efficiency, incorporating each lens's material transmission, thickness and coating efficacy, is represented by the purple solid curve in Figure~\ref{fig:throughput}.
Complete specifications for all optical components can be found in Table \ref{tab:LensData}.

%The optical design and optimization processes were meticulously carried out using ZEMAX software. 

The manufacturing tolerance parameters were guided by the {\sc OPTIMAX} precise tolerance level  \footnote{https://www.optimaxsi.com/charts/manufacturing-tolerance-chart/}. Results from tolerance analysis, including assembly tolerance errors, indicate that the spot RMS will remain below $3\,\mu$m for the 90th percentile across all three observing modes, based on dedicated Monte Carlo simulations.

\begin{table}[ht]
\caption{Collimator and camera lens prescription.} 
\label{tab:LensData}
\begin{center}       
\begin{tabular}{|l|c|c|c|c|c|c|}
\hline
Part & Surface & Material & Manufacturer & \makecell{Radius of \\ Curvature\\ $[$mm]} & Thickness & \makecell{Mechanical \\ Diameter\\ $[$mm]} \\  \hline
\hline
Collimator 1 (mirror)    & 1  & Fused Silica & CDGM   & 599.5   & --    & 108      \\ \hline
Collimator 2 (corrector) & 1  & H-K9L        & CDGM   & 2452.13 & 10    & 109      \\ \hline
Collimator 2 (corrector) & 2  & --           &  --    & 2538.13 & --    & 109      \\ \hline 
\hline
Camera 1                 & 1  & H-FK71       & CDGM   & 251.12  & 12    & 113      \\ \hline 
Camera 1                 & 2  & --           & --     & Inf     & --    & 113      \\ \hline 
Camera 2+3               & 1  & F4           & CDGM   & 217.25  & 10    & 83       \\ \hline   
Camera 2+3               & 2  & H-LAK52      & CDGM   & 97.83   & 10    & 83       \\ \hline   
Camera 2+3               & 3  & --           & --     & 406.53  & --    & 83       \\ \hline   
Camera 4                 & 1  & QF50GTI      & CDGM   & 131.86  & 8     & 63       \\ \hline   
Camera 4                 & 2  & --           & --     & 76.59   & --    & 63       \\ \hline   
 
\hline
\end{tabular}
\end{center}
\end{table}

% -----------------------------------------------------
\subsection{Disperser -- Ion-Etched Gratings}
% -----------------------------------------------------
\label{sec:grating}

We use novel binary mask gratings, ion-etched on a fused silica substrate, manufactured by {\sc Fraunhofer IOF, Jena}. 
This technique has already been proven successful for the UVVIS Son-of-X-Shooter (SOXS) spectrograph \cite{rubin2018mits}.
The gratings operate in Littrow configuration (first order -- $m=1$, keeping the anamorphic magnification factor r=1), with an incident angle of $45^\circ$.

The gratings have a ruled rectangle aperture with a width of $W= 134\, $mm. Due to the $45^\circ$ working angle and the $ d_1=100\,$mm beam diameter, the beam slightly overfills the grating, resulting in a loss of about $\sim 1.5 \% $ (assuming a uniformly illuminated footprint). A description of each grating, including line densities and simulated efficiencies, is summarized in Table~\ref{tab:gratingsParameters}, and the general properties of the gratings are summarized in Table~\ref{tab:generalGratingsParameters}.

%We are awaiting the final unit cell design parameters from the manufacturer. As shown in Fig. \ref{fig:Transmission}, two efficiency curves are presented for each grating, corresponding to different unit cell profile designs with the same groove density. While deeper etched grooves increase efficiency, they are more challenging to manufacture. Fraunhofer will attempt to produce the higher efficiency profiles, resorting to a more robust design with slightly lower efficiency if necessary.

%maybe discuss H alpha - FSR

\begin{table}[ht]
\caption{Parameters of the three ion-etched gratings. Each grating is optimized for its central wavelength with a narrow bandwidth. The gratings are manufactured by using atomic layer deposition on fused silica, allowing high line density with unprecedented precision.} 
\label{tab:gratingsParameters}
\begin{center}       
\begin{tabular}{|l|c|c|c|c|c|}
\hline
Line & $\lambda_\textrm{Littrow}$  & Bandpass Cover & Grooves Period $\sigma$ & Line Density & Grating Efficiency \\
\, & [nm]  & $\Delta \lambda$ [nm] &[grooves/mm] & [nm] & [\%] \\\hline
\hline
Ca\,II H\&K  & $395$  & 10.5 &$279.3$ & $3584.2$  & 95.1\\ \hline
Mg\,b       & $518$  & 13.8 &$365.6$ & $2724.8$  & 94.8\\  \hline
H$\alpha$   & $656$  & 17.5 &$463.9$ & $2155.2$  & 92.5\\  \hline
\end{tabular}
\end{center}
\end{table}

\begin{table}[ht]
\caption{Grating mechanical parameters} 
\label{tab:generalGratingsParameters}
\begin{center}       
\begin{tabular}{|l|c|c|c|c|c|}
\hline
Parameter & Specification           \\ \hline
\hline
Material & Fused Silica             \\ \hline
Grating ruled width, W & $134\,$mm   \\ \hline
Grating ruled length, L & $107\,$mm  \\ \hline
Substrate thickness & $6.35\,$mm     \\ \hline
\end{tabular}
\end{center}
\end{table}

% -----------------------------------------------------
\subsection{Detector -- Electron Multiplying CCD (EMCCD)}
% -----------------------------------------------------

When operating at high resolution with multiple traces, each from a small aperture telescope, the detector's readout noise (RO) of a typical CCD or CMOS detector becomes a significant noise contributor, as it is multiplied by the number of telescopes (traces).
To address this limitation, HighSpec employs an EMCCD -- Newton 971 unit from {\sc ANDOR Technology}. This CCD has a unique electron-multiplying structure built into the chip that amplifies the signal prior to readout, effectively bypassing the readout noise. The use of an EMCCD detector thus enhances our sensitivity, enabling the detection and quantification of single-photon events \cite{tulloch2011use}.

The detector's characteristics are detailed in Table \ref{tab:detector_parameters}. It provides high quantum efficiency (QE) across the Mg\,b and H$\alpha$ modes. The QE for the Ca\,II H\&K mode is more moderate and shows variation within the band, as illustrated in Figure \ref{fig:throughput}.

\begin{table}[ht]
\caption{EMCCD parameters} 
\label{tab:detector_parameters}
\begin{center}       
\begin{tabular}{|l|c|}
\hline
Image area (with 100\% fill factor) & 25.6 x 6.4 mm \\
\hline
Pixel size [$\mu \rm m$] & 16  \\
\hline
Pixel format & 1600$\times$400  \\
\hline
Dark current @ -100C [$\rm e^{-1}\ \rm s^{-1}\ \rm pixel^{-1}$]  & 0.0002 \\
\hline
Read noise [$\rm e^{-1}\rm pixel^{-1}$] & $\textless1$ \\
\hline
Peak quantum efficiency [\%] & $\textgreater95$ \\
\hline
Well capacity [$e^{-}$] & 1,300,000 \\ 
\hline
System window type & BV \\
\hline
\end{tabular}
\end{center}
\end{table}

% -----------------------------------------------------
\subsection{Opto-Mechanical Design}
% -----------------------------------------------------

A dedicated instrument room, with a controlled temperature to within $1^\circ$C, will house MAST instruments. The two spectrographs, HighSpec and DeepSpec, will be assembled together on an optical table equipped with active vibration isolation to ensure optimal stability.

This environment enables a streamlined mechanical design for the HighSpec spectrograph. All the optical elements are bonded to Al-6061 mounts using Dowsil 3145 RTV, simplifying construction. The Schmidt collimator optics, including a mirror and a corrector, are mounted on a separate base plate. The camera optics are mounted within a single cylindrical housing. These units are attached to the optical table through semi-kinematic interfaces.

The three gratings, corresponding to the spectrograph three observing modes, are positioned on a linear array mounted on a motorized stage \footnote{LRT0500AL-E08CT3A by {\sc Zaber}}. This setup allows for quick switching between gratings to change the observing mode. Each grating has its own holder within the frame to allow fine adjustments. A fourth frame slot has been included to accommodate additional gratings, offering the flexibility to expand observing capabilities in the future. The detector is separately mounted on a linear motorized stage \footnote{LRM025A-E03T3A by {\sc Zaber}}, enabling the designed focus adjustments for each observing mode (see section~\ref{sec:opticalDesign}).

\begin{figure} [h!]
   \begin{center}
   \begin{tabular}{c} %% tabular useful for creating an array of images 
   \includegraphics[height=7.5cm]{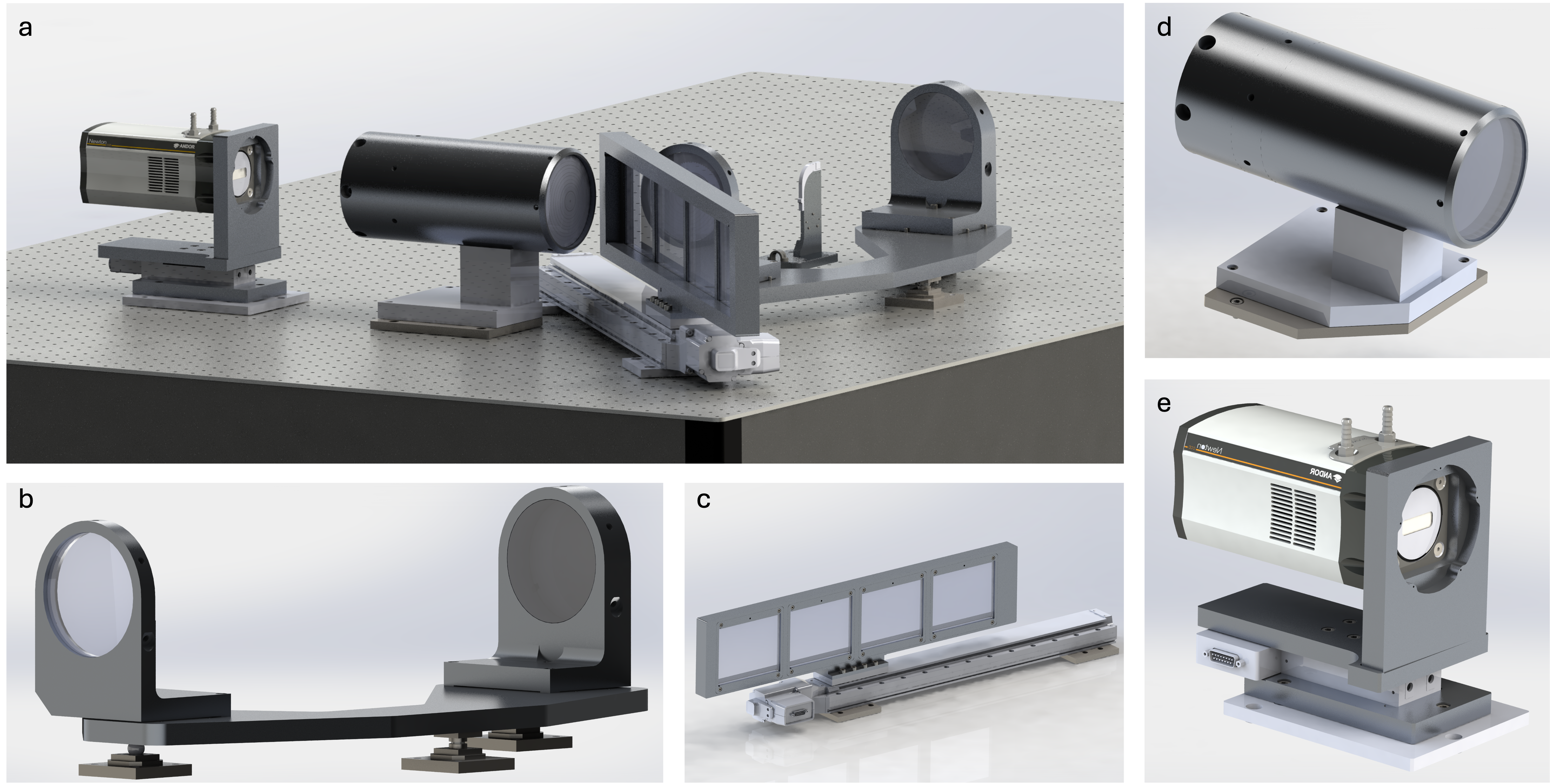}
   \end{tabular}
   \end{center}
   \caption[example] 
   { \label{fig:mechanicalMounting} HighSpec mechanical design. (a) Full optical path assembly. (b) Collimator (c) Gratings (d) Camera (e) Detector}
\end{figure}

Figure~\ref{fig:mechanicalMounting} illustrates the individual components of this mechanical design, showcasing the integration of each element.

%\vspace{5.5pt}

% -----------------------------------------------------
\subsection{Expected Performance}
\begin{figure} [h!]
   \begin{center}
   \begin{tabular}{c} %% tabular useful for creating an array of images 
   \includegraphics[height=12cm]{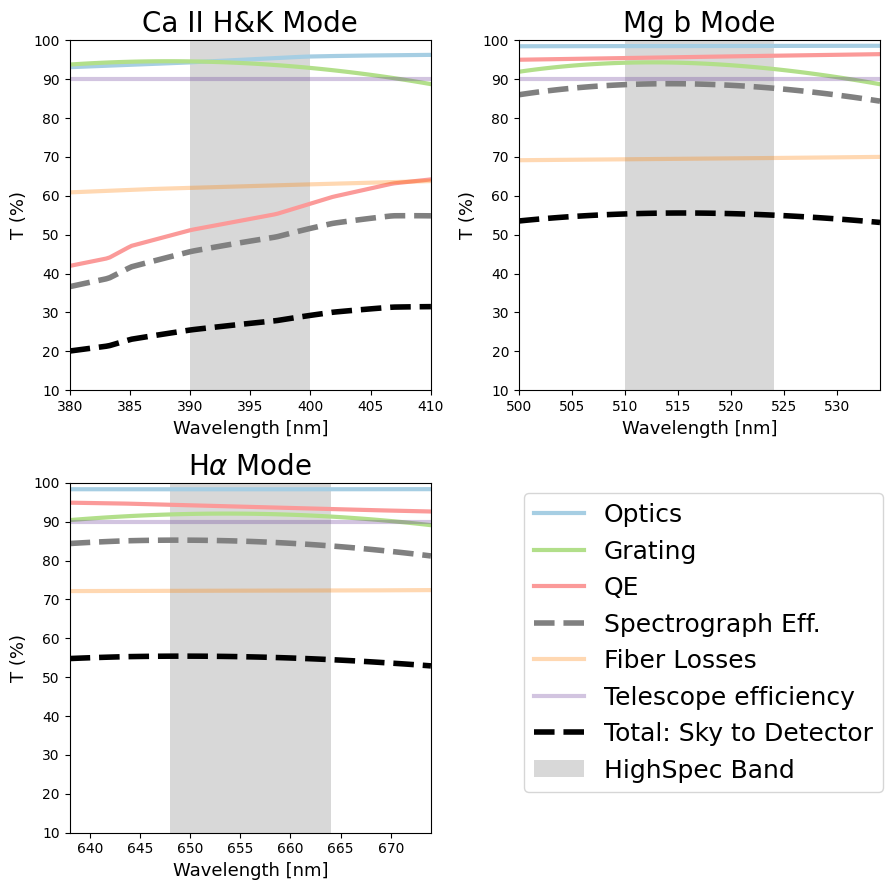}
   \end{tabular}
   \end{center}
   \caption[example] 
%>>>> use \label inside caption to get Fig. number with \ref{}
   { \label{fig:throughput} System Throughput. The gray dashed line represents the instrument's throughput -- slit to detector, incorporating all optics (both glass transmission and coatings losses), grating efficiency, and detector QE. The black dashed line illustrates the total expected efficiency -- sky to detector, accounting for additional losses from the telescope and fibers.}
\end{figure}

% -----------------------------------------------------
In Figure \ref{fig:throughput}, we present the expected performance of HighSpec's three observing modes. The figure highlights the slit-to-detector efficiency, with values exceeding 85\% for the Mg\,b and H$\alpha$ modes and over 55\% for the Ca\,II H\&K mode, largely due to the detector QE below $\sim400\,$nm. These efficiencies position HighSpec among the most effective spectrographs in its class. Additionally, the figure displays the total throughput for each mode, factoring in telescope reflection losses (estimated at 10\,\%), and fiber-related losses. These include simulated fiber transmission losses at the blue end (assuming 15 m long fibers), reflection losses at the fiber face (8\%), and losses due to focal ratio degradation (FRD; 6\%). Consequently, the total sky-to-detector efficiency amounts to $\gtrsim55\,\%$ for the Mg b and H$\alpha$ modes and $\gtrsim 31\%$ for the Ca II H\&K mode.

\begin{figure}[ht]
   \begin{center}
   \begin{tabular}{c} %% tabular useful for creating an array of images 
   \includegraphics[height=7cm]{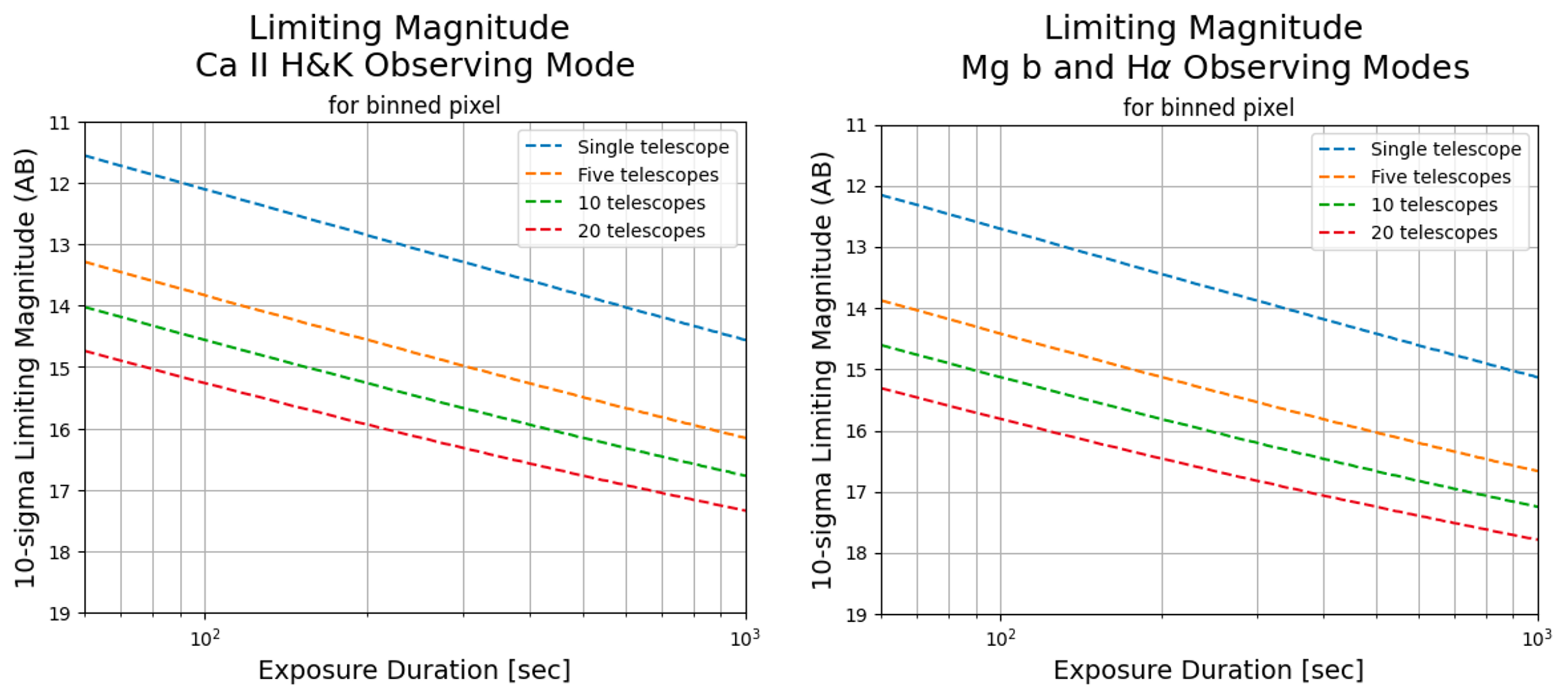}
   \end{tabular}
   \end{center}
   \caption[example] 
%>>>> use \label inside caption to get Fig. number with \ref{}
   { \label{fig:limitingMag}  10-$\sigma$ limiting magnitude of MAST+HighSpec as a function of exposure duration for different telescope configurations. Left Panel: The Ca\,II H\&K  observing mode. Right Panel: The H$\alpha$ observing mode. The main factor in the difference is the detector QE.}
\end{figure}

We have also simulated the 10$\,\sigma$ limiting magnitude for each observing mode under various telescope group configurations, see Figure~\ref{fig:limitingMag}. Noise factors are calculated independently for each imaged trace, incorporating the system’s expected throughput (as shown in Figure \ref{fig:throughput}), detector noise parameters (detailed in Table \ref{tab:detector_parameters}), an average sky brightness of AB mag = 20.5 \cite{ofek2023large}, and an atmospheric transmissivity of 85\,\%. The signal-to-noise ratio (SNR) was computed for a binned pixel (3 by 1 pixels, binning on the spatial direction) aligned with the dictated sampling resolution of HighSpec, which is set to exceed the Nyquist criteria covering three pixels (see Section~\ref{sec:opticalDesign}). When calculating the SNR per resolution element, the value increases by about a factor of $\sim\sqrt 3$.

\newpage

%\vspace{10pt}

% -----------------------------------------------------
\section{Current status}
% -----------------------------------------------------
HighSpec optical elements have been produced and received. The mechanical design is finalized, with mount fabrication set to be completed by the end of June 2024. Models were updated according to as-built data and assembly will commence in July. The gratings are currently being manufactured and are expected to be delivered in Q4 2024. The detector has been delivered, and dedicated testing is set to commence soon in our laboratory.  HighSpec is expected to be coupled with 10 telescopes by the end of 2024 and with the complete 20-telescope array by summer 2025.

% -----------------------------------------------------
\section{Science With HighSpec: The Study of White Dwarfs}
% -----------------------------------------------------
%%%%%%% intro about wds:

White dwarfs (WDs) are the most common stellar remnants, being the final evolutionary stage of low to inter-mediate-mass stars ($\lesssim$ 8-10 $M_\odot$), which represent over 97\,\% of the stars in the Milky Way. These compact remnants, roughly the size of Earth but with masses comparable to that of the Sun, are supported by electron degeneracy pressure and no longer sustain nuclear fusion in their cores and thus cool down for the remainder of their lifetimes.

WDs are distinguished by their high surface gravity, typically $g\approx10^8$\,cm\,s$^{-2}$. Due to this strong surface gravity, all elements heavier than hydrogen rapidly sink below the photosphere in timescale orders of magnitudes smaller than the cooling age of the star \cite{paquette1986diffusion,wyatt2014stochastic}. This gravitational separation leads to the predominance of pure hydrogen atmospheres in about 80\,\% of WDs, known as DA WDs. Others, with no hydrogen that survived the late phases of stellar evolution, will have a thin opaque helium layer -- the lightest element remaining, and are classified as DB WDs. Below a critical temperature, their optical spectra become featureless as they lack any absorption lines, leading to their classification as DC WDs.

HighSpec, specifically designed for the unique challenges of spectroscopic studies of WDs with moderate telescopes, features three observing modes optimized for in-depth studies of these stars. These modes facilitate targeted investigations of atmospheric contamination and detection of double-WD (DWD) systems by RV measurements.

%Observational Strategy and Sample

\begin{figure} [ht]
    \centering
    \includegraphics[height=7cm]{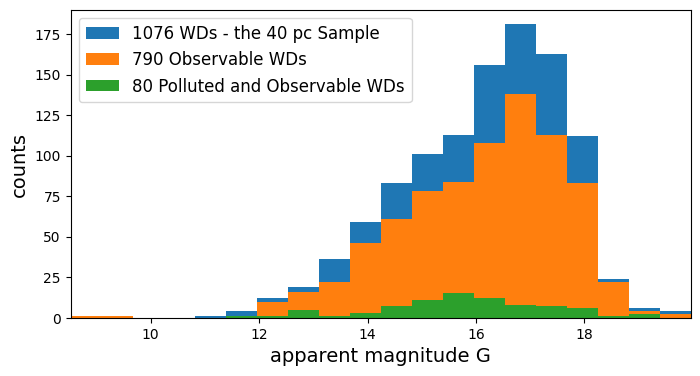}
    \caption{The 40\,pc WD sample based on Gaia data (blue), including the observable subset from Ne'ot Semadar (orange), and the subsample of metal-polluted WDs (green). Observability is defined as any target that can be observed from Ne'ot Smadar for more than three hours per year with an airmass below 2.} 
    \label{fig:40pc_sample} 
\end{figure}

Our initial observational efforts with HighSpec will concentrate on the 40\,pc WD sample \cite{o202440}, which is the largest unbiased volume-limited sample of WDs \footnote{The coolest DZ WD known to date, WD\,J2147$-$4035 \cite{elms2022spectral}, has an absolute magnitude of $M_{G}$ = 17.73 -- implying it would not be detected by Gaia at distances larger than $\approx40$\,pc \cite{tremblay2024gaia}}. 
This sample includes 1076 spectroscopically confirmed WDs, providing near-complete coverage with 99.3\% spectroscopic completeness.  In addition, the completeness of the Gaia DR3 WD selection from Gentile-Fusillo et al., (2021) catalog 
 \cite{gentile2021catalogue} at 40\,pc is estimated to be $\approx97\,\%$ based on pre-Gaia surveys and population synthesis \cite{toonen2017binarity,hollands2018gaia,mccleery2020gaia}.
This extensive survey will allow us to examine a comprehensive and representative group of local WDs, enhancing the reliability of our findings. From the Ne'ot Smadar observatory, 790 WDs are observable, of which 80 are already identified as contaminated by heavy elements \cite{o202440}, see Figure~\ref{fig:40pc_sample}.

% -----------------------------------------------------
\subsection{Detection of Polluted WDs from the Ca\,II H\&K and Mg\,b Absorption Lines}
% -----------------------------------------------------
% intro
In the absence of physical mechanisms competing with the gravitational settling, the optical spectra of WDs typically exhibit lines corresponding only to their atmospheric composition: hydrogen lines (DA), helium lines (DB), or pure continuum in WDs much cooler the H/He ionization temperature (DC). However, some WDs display traces of heavy elements in their atmosphere (e.g., Na, Mg, Si, Ca, Fe), indicating the influence of additional processes beyond gravitational settling. 
For most of these "polluted" WDs, atmospheric phenomena such as dredge-up  \cite{koester1982atmospheric} and radiative levitation \cite{chayer1995radiative} are insufficient to explain the observed metallic content, suggesting an external origin for these materials. 

Typically, these pollutants are rocky and non-volatile, with compositions closely resembling Earth's bulk material \cite{hollands2018cool,doyle2020extrasolar}. The accumulated mass of these pollutants generally reaches up to about $10^{-6} M_\oplus$ \cite{xu2012spitzer}, indicating that they likely originate from disintegrated rocky bodies similar to asteroids or minor planets within the WD's system.

% interpretation:
The current interpretation posits that these pollutants come from surviving planetesimals, similar to asteroids in our solar system. In this scenario, planets that have survived the red-giant phase may occasionally perturb these asteroids, causing them to scatter inward from several AUs to the tidal zone of the white dwarf, where they settle into concentric orbits and form debris disks. Dynamical instabilities within these disks then disrupt the material, creating a dust disk that eventually accretes onto the WD \cite{jura2014extrasolar}. This theory is supported by various observational methods, including near-infrared excesses indicative of dust disks, double-peaked optical emission features from gas disks, and X-ray emissions from ongoing accretion processes \cite{veras2021planetary, malamud2024white}.

% why is it interesting:
Detecting and analyzing these signatures in WD atmospheres can provide insights into the composition of extrasolar planetary materials, as the accreted matter directly probes exoplanetary chemistry. Additionally, it can shed light on the occurrence rates of planetary systems around massive stars -- a demographic poorly understood due to observational biases favoring lower-mass stars. Furthermore, detecting and accurately modeling the pollutants in WDs are crucial for deriving the precise WD parameters, especially for cool WDs \cite{blouin2020magnesium}.

%Current Observations of Polluted White Dwarfs: Limitations and Findings
The detection of metal pollution in WDs is influenced by several factors, including the WD's characteristics (such as age, mass, and distance) and the capabilities of the observational instruments (mainly resolution and sensitivity). For instance, Zuckerman et al. (2003) \cite{zuckerman2003metal} observed a sample of single DA WDs with effective temperatures below 10,000\,K using high-resolution visible spectra from Keck HIRES and found that approximately 25\% exhibited metal pollution. In contrast, the comprehensive 40\,pc WD sample \cite{o202440}, including spectra of varying resolution and SNR (mostly SDSS), indicates a pollution rate of only 11\% for WDs with comparable properties \footnote{O'Brien et al. (2023) using the X-shooter instrument reported a metal pollution rate of about 15\% within a subsample of WDs in the 40 pc under similar sample conditions}.

%Zuckerman et al. (2003) \cite{zuckerman2003metal} observed that approximately 25\% of single DA WDs with effective temperatures below 10,000 K exhibited metal pollution when using high-resolution visible spectra from Keck HIRES. In contrast, O'Brien et al. (2023) reported a metal pollution rate of about 15\% within a subsample of WDs in the 40 pc using the X-shooter instrument, under similar sample conditions. Additionally, the comprehensive 40 pc WD sample \cite{o202440} indicates a pollution rate of only 9\% for WDs with comparable properties.

% I can summarize this paragraph to one sentance
These findings suggest that the extant data on metal-polluted WDs within the 40\,pc sample remain limited and vary significantly depending on the observational properties. Many WDs in the 40 pc sample still need to be analyzed with high-resolution, high signal-to-noise ratio (SNR) spectroscopy, as most documented cases of metal pollution are based on low to medium-resolution optical spectra. Obtaining higher resolution and quality spectra for these WDs is crucial for accurately updating the fractions of metal-polluted WDs, determining material abundance, and understanding the underlying distributions within this volume-limited sample.

% the exact work with HighSpec
HighSpec's observing modes will allow us to study polluted WDs. The Ca II H\&K lines ($3968.5,\text{\AA}$ and $3933.7,\text{\AA}$) are reliable indicators of pollution as Ca is the most detectable chemical species in the optical spectrum \cite{zuckerman2003metal, coutu2019analysis}. Magnesium is the second most abundant element in polluted WDs \cite{coutu2019analysis}. The combination of the two observing modes will allow us to better determine the Mg/Ca ratio, essential for understanding the origin of the accreted material \cite{blouin2020magnesium}. In addition, comparing the RVs measured from the Ca\,II lines and H$\alpha$ can help identifying whether these metal lines are of photospheric origin indeed (and not from the interstellar medium). 

%HighSpec will survey all observable WDs in the 40 pc sample from Ne'ot Smadar, focusing on DA WDs initially, as this subset is also targeted for the RV survey. Focusing on specific WDs simplifies telescope array operations since the switching between HighSpec's observing modes is motorized and easily controlled.

% what do we exactly suggest:
%Therefore, HighSpec will survey all observable WDs in the 40\,pc sample from Ne'ot Smadar, utilizing a dedicated observing mode centered on the Ca II H\&K and Mg b triplet lines. Initially, the survey will concentrate on DA WDs as this particular subset is also the target for the RV survey. Focusing on specific WDs each time simplifies the telescope array operations, as switching between observing modes in HighSpec is motorized and easily controlled.

%This extensive sample will allow us to focus on a complete and representative group of local WDs, enhancing the reliability and relevance of our findings. From the observatory in Ne'ot Smadar, 790 WDs are observable, from which 80 are already identified as contaminated by heavy elements based on spectral analysis, see Figure~\ref{fig:40pc_sample}. 

% -----------------------------------------------------
\subsection{Statistical Analysis of the Local Close-Orbit Double White Dwarf Population: Radial Velocity Measurements  from the $H_{\alpha}$ Absorption Line}
\label{sec:RV}
% -----------------------------------------------------

The population of close-orbit double WDs (DWDs) might hold the clue for several astrophysical phenomena -- from binary stellar evolution, through Type Ia supernova (SN Ia) progenitors, to gravitational-wave sources. 
Despite its importance, our current knowledge of the DWDs population --- its frequencies, period distribution, and component-mass distribution -- is rudimentary at best \cite{maoz2018separation}. 

The high surface gravity of WDs usually leaves them with only a few hydrogen absorption lines in their spectra, with a typical width of $\sim50$\,\AA\ -- posing a great challenge for high-precision radial-velocity (RV) measurements \cite{rebassa2019double}.
The most precise RV measurements of WDs have utilized the well-resolved core of the H$\alpha$ line with a typical width of $\sim1.5$\,\AA, caused by non-local thermal equilibrium (NLTE) effects.
By fitting a Gaussian profile to the NLTE cores, using a nonlinear least-squares fitting routine \cite{falcon2010gravitational, maoz2017}, these measurements can achieve a precision of a few km/s, corresponding to DWDs in orbits with up to a few AUs.

As higher-order Balmer lines usually do not have a visible NLTE core, a spectroscopic measurement of the narrow band surrounding the H$\alpha$ line, with a resolution capable to resolve to NTLE line core, is sufficient for these WDs RV characterization. Maoz et al. (2012) showed that with two RV epochs per WD, the $\Delta$RV distribution can set meaningful constraints on the binary fraction of the population, $f_\text{bin}$, and on the distribution of binary separations, $dN/da$ \cite{maoz2012characterizing}. Statistical inference about the DWD population is thus possible with limited follow-up observations and without full binary orbital solutions for the candidates.

The Supernova Ia Progenitor surveY (SPY) sample, obtained with UVES on the VLT in the early 2000's, is still the largest multi-epoch high-resolution spectroscopic sample of WDs \cite{napiwotzki2020eso}. It includes about 650 DA WDs brighter than $B\sim 16.5$\,mag with multi-epoch RV measurements. However, it inherited the significant and largely unclear selection biases of the small sample of WDs known at that time. %However, using a sub-sample of 439 WDs, \cite{maoz2017binary} constrained the DWDs fraction and merge rate to be suffice to produce most or all SNe Ia.

HighSpec offers a dedicated observing mode centered around the H$\alpha$ line. The expected efficiency of the spectrograph is 85\,\%, and the end-to-end efficiency is 55\,\%, allowing to observe WDs up to a limiting magnitude of 17.5 (see Figure~\ref{fig:limitingMag}, right panel). 

Gaia DR2, released in 2018, provided for the first time an unprecedentedly large and unbiased sample of WDs across the entire sky. According to its successor, the Gaia EDR3 WDs catalog \cite{gentile2021catalogue}, HighSpec will be able to obtain multi-epoch, high-resolution spectroscopy for about 6,500 DA white dwarfs\footnote{This includes a wide range of hybrid spectral types, such as DAB and DAZ, not solely DA types.} --- about ten times the SPY sample size. This includes 94\,\% (479 WDs) of the DA 40\,pc sample of WDs in the northern hemisphere \cite{o202440}, in addition to many other WDs. The selection biases will be easy to control and correct using complementary Gaia data.% In addition, we will be able to conduct follow-up RV observations on transiting WDs identified by the MAST photometry observations that will allow us to better constrain the nature of the variability in their light curve.
Thus, HighSpec will provide for the first time, a large controlled high-resolution spectroscopic sample of WDs, deepening our knowledge of the poorly understood process of binary evolution.

\acknowledgments % equivalent to \section*{ACKNOWLEDGMENTS}       
 
S.B.A. is grateful for support from the Azrieli Foundation, André Deloro Institute for Advanced Research in Space and Optics, Peter and Patricia Gruber Award, Willner Family Leadership Institute for the Weizmann Institute of Science, Aryeh and Ido Dissentshik Career Development Chair, Israel Science Foundation, Israel Ministry of Science, and Minerva Stiftung.

% References
%\bibliography{report} % bibliography data in report.bib
%\bibliographystyle{spiebib} % makes bibtex use spiebib.bst

\end{document}